\documentclass[12pt,a4paper]{article}
\usepackage{graphicx}
\renewcommand{\baselinestretch}{2}
\begin{document}
\title{Influence of electric fields on absorption spectra of AAB-stacked trilayer graphene\\}
\author{Chih-Wei Chiu$^{1,a}$  and Rong-Bin Chen$^{2,b}$\\
{\small $^{1}$Department of Physics, National Kaohsiung Normal University, 824 Kaohsiung, Taiwan}\\
{\small $^{2}$Center of General Studies, National Kaohsiung Marine University, 811 Kaohsiung, Taiwan}}
\date{\today}
\maketitle

\begin{abstract}

The band structures and optical properties of AAB-stacked trilayer graphenes (AAB-TLG) are calculated by the tight-binding model and gradient approximation. Three pairs of  the energy bands exhibit very different energy dispersions at low energy and saddle points at the middle energy. At zero electric field, $3$$^2$ excitation channels exist in both the low and middle frequencies, and cause the very rich joint density of states (JDOS). However, the structures in the JDOS do not appear in the absorption spectra completely. Due to the different contributions from the velocity metric elements, some excitation transitions disappear in the spectra. Furthermore, the frequency and the existence of the absorption structures are affected by the increase of the electric field from zero.

\vskip 1.0 truecm $^{a)}$ Electronic mail: giorgio@fonran.com.tw
\vskip 0.0 truecm $^{b)}$ Electronic mail: rbchen@mail.nkmu.edu.tw
\pagebreak

\pagebreak
\renewcommand{\baselinestretch}{2}
\end{abstract}
\renewcommand{\baselinestretch}{2}

A two-dimensional (2D) graphene layer has very strong $sp^2$ bonding. The $\pi$ bands made up from $2p_z$ orbitals determine many of the physical properties. In recent years, few-layer graphenes, a stack of graphene layers with thickness less than 50 nm, have been produced by mechanical exfoliation of highly oriented pyrolytic graphite [1] and epitaxial growth on silicon carbide [2]. They are semimetals with slight overlapping near the Fermi level between the conduction and the valence bands. The band structures can be changed by stress [3]. According to the stacking sequence, there are three graphene crystals: [4-8] (1) hexagonal simple graphene with AA stacking, (2) Bernal graphene with AB stacking, and (3) rhombohedral graphene with ABC stacking. As a result of the
hexagonal symmetry and stacking configurations, these systems exhibit rich physical properties. Theoretical approaches have been used in the investigations of magnetic properties [9], quantum Hall effect [9,10], phonon spectra [10], electronic properties [11,12], and optical excitations [13,14]. Accordingly, these materials are considered highly promising for applications in future device [15-19]. Recently, the new system with the AAB stacking start to be investigated, and this stacking is the last of the possible configurations for the trilayer systems [20,21].

We develop the generalized tight-binding model, based on the subenvelope functions on the distinct sublattices, to study the rich electronic properties of AAB-stacked trilayer graphene. This work shows that the AAB-stacked configuration presents an abnormal band structure, in which there are pairs of oscillatory, sombrero-shaped and parabolic bands, which are different from those of the other stacking systems mentioned above. The special band structure could be verified by angle-resolved photoemission spectroscopy (ARPES).

The low-energy $\pi$-electronic structure of AAB-stacked trilayer graphene (AAB TLG), mainly coming from the 2$p_z$ orbitals, is calculated with the generalized tight-binding model. The two sublattices in the $l$th ($l$=1, 2, 3) layer are denoted as A$^l$ and B$^l$. The first two layers, shown in Fig. 1, are arranged in the AA-stacking configuration; that is, all carbon atoms have the same (x,y) projections. The third layer can be obtained by shifting the first (or the second) layer by a distance of $b$ along the armchair direction. In this system, the A atoms (black) have the same ($x,y$) coordinates, while the B atoms (red) on the third layer are projected at the hexagonal centers of the other two layers. The interlayer distance and the C-C bond length are, respectively, $d$=3.37 $\AA$ and $b$=1.42 $\AA$. There are six carbon atoms in a primitive unit cell. The low-energy electronic properties are characterized by the carbon 2$p_z$ orbitals. The zero-field Hamiltonian, which is built from the six tight-binding functions of the 2$p_z$ orbitals, is dominated by the intralayer and the interlayer atomic interactions $\gamma$ is. There exist 10 kinds of atom-atom interactions corresponding to the 10 atomic hopping integrals which appear in the Hamiltonian matrix. $\gamma_0$=2.569 eV represents the nearest-neighbor intralayer atomic interaction; $\gamma_1$=0.263 eV, $\gamma_2$=0.32 eV present the interlayer atomic interactions between the first and second layer; $\gamma_3$=0.413 eV, $\gamma_4$=0.177 eV, $\gamma_5$=0.119 eV are associated with the interlayer atomic interactions between the second and third layer; $\gamma_6$=0.013 eV, $\gamma_7$=0.0177 eV, and $\gamma_8$=0.0119 eV relate to the interlayer atomic interactions between the first and third layer; and $\gamma_9$=0.012 eV accounts for the difference in the chemical environment of A and B atoms. The hopping integrals $\gamma_1$, $\gamma_3$, and $\gamma_6$ belong to the vertical interlayer atomic interactions, while the others are non-vertical ones.

The zero-field band structure of AAB TLG consists of three pairs of conduction and valence subbands dispersed in the energy-wave-vector space along $\Gamma$ $\to$ K $\to$ M $\to$ $\Gamma$.
The pairs according to the magnitude of the K state energy from small to large are labeled $E^{c_1,v_1}$, $E^{c_2,v_2}$, and $E^{c_3,v_3}$, as shown by the black curves in Fig. 2.
Near the Fermi energy (the inset), the two subbands which belong to the first pair, $E^{c_1,v_1}$, have strong oscillatory energy dispersions.
The conduction subband starts to increase from $E^{c_1}_{K}$ (a local minimum value of about 4 meV at the K point), along the $\Gamma$K and KM directions.
After reaching $E^{c_1}_{b_U}$ (a local maximum value of about 58 meV at the band edge state of $b_U$), it decreases until reaching $E^{c_1}_{b_D}$ (a local minimum energy of about 4 meV at the band edge state of $b_D$), and then grows steadily.
The curvature of the valence subband is in the opposite direction, which is almost symmetric to the conduction subband about $E_F$.
The first pair of subbands, with three constant energy contours within $\pm$58 meV and a narrow gap $E_g\sim$ 8 meV in between them, is special in that it has never appeared in any other stacking configuration.
The second pair of subbands, $c_2 \& v_2$, has a sombrero-shaped and a local energy minimum (maximum) and maximum (minimum), situated at around 0.24 (-0.24) eV and 0.26 (-0.26) eV, respectively.
The energy difference between the two extreme points is quite narrow, being only about 20 meV. Located away from $E_F$, the third pair of subbands, $c_3 \& v_3$, consists of monotonic parabolic bands with a minimum (maximum) value of about 0.49 (-0.49) eV.
For the middle energy, the states at the $M$ point are saddle points, i.e. the state energy is a maximum (minimum) along $K \rightarrow M$ ($M \rightarrow \Gamma$). Therefore, there exists the larger density of states at the energy of the M points ($E^{c,v}_{M}$).
When a perpendicular electric field F (expressed in units of V/$\AA$) is applied along the stacking direction, the band structures exhibit huge variations.
As F grows from zero, the influences are as follows. For the low energy, the $b_M$ and $b_m$ states become far from the K point, and the values of $|E^{c_1,v_1}_{b_U}|$ increase. Both $|E^{c_1,v_1}_{K}|$ and $|E^{c_1,v_1}_{b_U}|$ rise and then reduce, and the former has a especially obvious change. Furthermore, $|E^{v_2}_{K}|$ decreases and then increases, but $|E^{c_2}_{K}|$ and $|E^{c_3,v_3}_{K}|$ increase monotonically.
As for the middle energy, the first-pair subbands close to (distant from) the low energy, that is $|E^{c_1,v_1}_M|$ ($|E^{c_3,v_3}_M|$) reduces (enlarges). $E^{c_2}_M$ decreases and then increases slightly, and $E^{v_2}_M$ does not show a clear change.


At zero temperature, electrons in AAB TLG are assumed to be excited by the EM field with polarization $\hat{\bf E}$ from the occupied states to the unoccupied states. The spectral function obtained from Fermi$'$s golden rule is given by
\begin{eqnarray}
{\rm A(\omega)}&\propto &\sum_{{h^\prime},h}{\int_{1stBZ}} {d^3{\bf k}\over\,(2\pi)^3} {|\langle\Phi^{h^\prime}({\bf k})|\hat{\bf E}\cdot {\bf P}/m_{e}| \Phi^h({\bf k})\rangle|^2} \cr
&&\times\frac{\delta[f(E^h({\bf
k}))-f(E^{h\prime}({\bf k}))]} {(E^{h^\prime}({\bf k})-E^h({\bf
k})-\omega)^2+\delta^2},
\end{eqnarray}
where $\Phi^{h^\prime}({\bf k})$ and  $\Phi^{h}({\bf k})$ [$E^{h^\prime}({\bf k})$ and $E^{h}({\bf k})$] are the wave
functions [state energies] of the final and initial states, and $h$ ($h^\prime$) represents either conduction or valence bands. $\delta$= 0.01 eV is the energy width due to various deexcitation mechanisms and $f(E^h({\bf k}))$ is the Fermi-Dirac distribution function.
The dipole matrix element, ${\rm M}^{cv}({\bf k})={\langle\Phi^{h^\prime} ({\bf k})|\hat{\bf E}\cdot {\bf P}/m_{e}|\Phi^h({\bf k})\rangle}$, could be evaluated within the gradient approximation\cite{12}. It is anisotropic for $\hat{\bf E}$, but the integral of $|{\rm M}^{cv}({\bf k})|^2$ presents the isotropic property.


The joint density of states ($D_J$), which is defined by setting the velocity matrix elements $|{\rm V}^{h^{\prime},h}({\bf k})|$ in eq.(1) to one, is shown in Fig. 3 for the different electric fields. It is proportional to the number of optical excitation channels. When there exists a great number of states with very close excitation energy, $D_J$ exhibits a prominent structure. The 3$\times $3 excitation channels in both the low and middle frequency caused from the saddle points or band edge states of the three pairs of the energy bands lead to rich structures spreading out in $D_J$. The frequencies of those structures correspond to the energy spacings between the conduction bands and the valence bands at the specific states.
For the low frequency [the upper inset], a shoulder ($s_1$), a weak-peak ($s_2$), a peak structures ($s_3$) are respectively located at $\omega\sim$ 16 meV, $\omega\sim$ 0.1 eV, and $\omega\sim$ 0.32 eV corresponding to $E^{c_1}_{K}-E^{v_1}_{K}$ $\&$ $E^{c_1}_{b_D}-E^{v_1}_{b_D}$, $E^{c_1}_{b_U}-E^{v_1}_{b_U}$, and $E^{c_1}_{b_U}-E^{v_2}$ $\&$ $E^{c_2}-E^{v_1}_{b_U}$ at the zero field [black curve]. Furthermore, there are three broad peaks ($s_4$, $s_5$, $s_6$) at $\omega\sim$ (0.51 eV, 0.61 eV, 0.65 eV), associated with the transitions of ($E^{v_2}\to E^{c_2}$, $E^{v_1}\to E^{c_3}$, $E^{v_3}\to E^{c_1}$). Under the influence of the electric field, the forms and the frequencies of these structures exhibit eminent variations. Most structures are shifted to higher frequencies with the increase of F, and the $s_1$ structure, due to the split of $E^{c_1}_{b_D}-E^{v_1}_{b_D}$ $\&$ $E^{c_1}_{K}-E^{v_1}_{K}$, are separated to the s1 and s1$'$ ones.
As for the middle frequency, there are nine peaks ($p_1...p_9$) induced from the M points, and they correspond to ($v_1\to c_1$, $v_1\to c_2$, $v_2\to c_1$, $v_1\to c_3$, $v_2\to c_2$, $v_3\to c_1$, $v_2\to c_3$, $v_3\to c_2$, $v_3\to c_3$) [the lower inset]. The range of these peak frequencies becomes wider quickly as F grows, but the three peaks of $p_4, p_5, p_6$ in the center of the nine ones show a smaller shift. Some peaks with the same frequency overlap at the certain F, such as $p_7$ and $p_8$ at F=0.2.
The high $D_J$ from the electronic structures might be reflected in absorption spectra.


The optical absorption spectra, A($\omega$), are shown for AAB TLG at different F$'$s in Fig. 4. Because of the different contributions of $|{\rm M}^{cv}({\bf k})|^2$ among the states, the features of $D_J$ may not be completely reflected in A$\omega$), e.g. the strength and the form of the structures. The spectra exhibit complex structures in the low (1.7 eV$>\omega$) and middle (6.5 eV$>\omega>$ 4.4 eV) frequency ranges; otherwise they are identical among different $F$$'$s. The former reflects the optical excitations in the vicinity of the $K$ points (or at the band edge states), and the latter is for the excitations at the $M$ points.
In the small frequency region (upper inset), there are several absorption structure at F=0, which can be find in $D_J$. When the electric field grows from zero, the smaller-frequency A($\omega$) exhibit stronger and stronger. For example, the first peak is the strongest one at F=0.2.
As to the middle frequency (lower inset), the nine excitation channels do not wholly cause obvious structures in A($\omega$). The $s_3$ and $s_7$ of the $D_J$ peaks disappear at zero F. With the increase of the electric field, the range of the peak frequency extend, and the peaks with the large shift are harder to survive. The peaks of $s_1$, $s_2$, and $s_9$ vanish at F=0.2,

In conclusion, it can be said that the tight-binding model and gradient approximation are, respectively, used to calculate the band structures and the absorption spectra of AAB-stacked trilayer graphenes. Three pairs of energy bands include a strong oscillatory, a sombrero-shaped, and a parabolic energy dispersions at the low energy, and form saddle points at the middle energy. At zero electric field, $3\times 3$ excitation channels exist in both the low and middle frequencies, and cause the very rich joint density of states (JDOS). However, the strong-JDOS transitions are not reflected in the absorption spectra completely. In the absorption spectra, because of the different strength of the velocity metric elements among the electron states, two middle-$\omega$ peaks which exist in the JDOS vanish at zero electric field. When the electric field along the stacking direction grows from zero, the forms of the structures are changes, and the frequencies enlarge at the low frequency. Furthermore, the vanished peaks reappear, and the other peaks vanish at middle frequency.

This work was supported in part by the National Center for Theoretical Sciences and the National Science Council of Taiwan, under Grant numbers NSC 98-2112-M-145-003-MY3,
NSC 101-2811-M-006-013-,
NSC 101-2112-M-244-001-, and
NSC 98-2112-M-006-013-MY4.

\newpage

{\Large\bf References}

\begin{itemize}

\item[{1}]
K. S. Novoselov, A. K. Geim, S. V. Morozov, D. Jiang, Y. Zhang, S.
V. Dubonos, I. V. Grigorieva and A. A. Firsov: Science \textbf{306}
(2004) 666.

\item[{2}]
C. Berger, Z. Song, X. Li, X. Wu, N. Brown, C. Naud, D. Mayou, T.
Li, J. Hass, A. N. Marchenkov, E. H. Conrad, P. N. First and W. A.
de Heer: Science \textbf{312} (2006) 1191.

\item[{3}]
Z. H. Ni, T. Yu, Y. H. Lu, Y. Y. Wang, Y. P. Feng and Z. X. Shen: ACS
Nano \textbf{2} (2008) 2301.

\item[{4}]
J. W. McClure: 1969 Carbon \textbf{7} (1969) 425.

\item[{5}]
Z. Y. Rong and P. Kuiper: Phys. Rev. B \textbf{48} (1993) 17427.

\item[{6}]
P. J. Ouseph: Phys. Rev. B \textbf{53} (1996) 9610.

\item[{7}]
J. C. Charlier, X. Gonze and J. P. Michenaud: Carbon \textbf{32}
(1994) 289.

\item[{8}]
J. C. Charlier, J. P. Michenaud and X. Gonze: Phys. Rev. B
\textbf{46} (1992) 4531.

\item[{9}]
E. McCann and V. I. Fal'ko: Phys. Rev. Lett. \textbf{96} (2006)
086805.

\item[{10}]
S. Kitipornchai, X. Q. He and K. M. Liew: Phys. Rev. B \textbf{72}
(2005) 075443.

\item[{11}]
Y. H. Lai, J. H. Ho, C. P. Chang and M. F. Lin: Phys. Rev. B
\textbf{77} (2008) 085426.

\item[{12}]
J. H. Ho, C. L. Lu, C. C. Hwang, C. P. Chang and M. F. Lin: Phys.
Rev. B \textbf{74} (2006) 085406.

\item[{13}]
C. L. Lu, C. P. Chang, Y. C. Huang, R. B. Chen and M. F. Lin: Phys.
Rev. B \textbf{73} (2006) 144427.

\item[{14}]
Y. H. Ho, Y. H. Chiu, D. H. Lin, C. P. Chang and M. F. Lin: ACS
Nano \textbf{4} (2010) 1465.

\item[{15}]
J. C. Charlier, X. Blase and S. Roche: Rev. Mod. Phys. $\textbf{79}$
(2007) 677.

\item[{16}]
A. K. Geim and K. S. Novoselov: Nat. Mater. $\textbf{6}$ (2007) 183.

\item[{17}]
Z. Wu, Z. Chen, X. Du, J. M. Logan, J. Sippel, M. Nikolou, K.
Kamaras, J. R. Reynolds, D. B. Tanner, A. F. Hebard and A. G.
Rinzler: Science \textbf{305} (2004) 1273.

\item[{18}]
V. Sazonova, Y. Yaish, H. Ustunel, D. Roundy, T. A. Arias and P. L.
McEuen: Nature \textbf{431} (2004) 284.

\item[{19}]
B. Bourlon, C. Miko, L. Forro, D. C. Glattli and A. Bachtold: Phys.
Rev. Lett. \textbf{93} (2004) 176806.

\item[{20}]
Thi-Nga Do, Chiun-Yan Lin, Yi-Ping Lin, Po-Hsin Shih, and Ming-Fa Lin
Carbon \textbf{94} (2015) 619-632.

\item[{21}]
arXiv:1509.02253v1

\end{itemize}

\newpage
{\Large\bf Figure Captions}\\
\begin{itemize}
\item[FIG. 1.] The interlayer atomic interactions and the geometric structure under a uniform electric field F$\hat z$. The shaded region corresponds to a rectangular unit cell. The first and second layers have the same (x, y) projections.

\item[FIG. 2.] Energy bands of an AAB-stacked trilayer graphene  are shown along different directions in the first Brillouin zone. The inset is detail results of $E^c_1,v_1$.

\item[FIG. 3.] The joint density of states for different F$'$s. The upper and lower insets are the detail results in the low and middle frequency, respectively.

\item[FIG. 4.] The optical absorption spectra for different F$'$s. The upper and lower insets are the detail results in the low and middle frequency, respectively.
    
\end{itemize}

\end{document}